\documentclass[journal,twoside,web]{ieeecolor}
\usepackage{tmi}
\usepackage{cite}
\usepackage{amsmath,amssymb,amsfonts}
\usepackage{algorithmic}
\usepackage{graphicx}
\usepackage{textcomp}
\def\BibTeX{{\rm B\kern-.05em{\sc i\kern-.025em b}\kern-.08em
    T\kern-.1667em\lower.7ex\hbox{E}\kern-.125emX}}
\begin{document}
\title{A Physics-based Generative Model to Synthesize Training Datasets for MRI-based Fat Quantification}
\author{Juan P. Meneses, Yasmeen George, Christoph Hagemeyer, Zhaolin Chen\textsuperscript{\textsection}, Sergio Uribe\textsuperscript{\textsection}
\thanks{Juan P. Meneses is with the
Department of Electrical Engineering, Pontiﬁcia Universidad Cat\'olica de Chile, Santiago 7820436, Chile, the Biomedical Imaging Center, Pontiﬁcia Universidad Cat\'olica de Chile, Santiago 7820436, Chile, and also with the Millennium Institute for Intelligent Healthcare Engineering, Santiago,
Chile. E-mail: jpmeneses@uc.cl.}
\thanks{Sergio Uribe is with the
Department of Medical Imaging and Radiation Sciences, Monash University, Melbourne, VIC, Australia. E-mail: sergio.uribe@monash.edu.}
\thanks{Yasmeen George is with the
Department of Data Science and AI, Monash University, Melbourne, VIC, Australia. E-mail: yasmeen.george@monash.edu.}
\thanks{Zhaolin Chen and Christoph Hagemeyer are with Monash Biomedical Imaging, Monash University, Melbourne, VIC 3168, Australia. E-mail: \{zhaolin.chen, christoph.hagemeyer\}@monash.edu.}
\thanks{This manuscript has been submitted to IEEE Transactions on Medical Imaging for possible publication on November 14, 2024. Copyright may be transferred without notice after acceptance.}}

\maketitle
\begingroup\renewcommand\thefootnote{\textsection}
\footnotetext{Equal contribution}
\endgroup

\begin{abstract}
Deep learning-based techniques have potential to optimize scan and post-processing times required for MRI-based fat quantification, but they are constrained by the lack of large training datasets. Generative models are a promising tool to perform data augmentation by synthesizing realistic datasets. However no previous methods have been specifically designed to generate datasets for quantitative MRI (q-MRI) tasks, where reference quantitative maps and large variability in scanning protocols are usually required. We propose a Physics-Informed Latent Diffusion Model (PI-LDM) to synthesize quantitative parameter maps jointly with customizable MR images by incorporating the signal generation model. We assessed the quality of PI-LDM's synthesized data using metrics such as the Fr\'echet Inception Distance (FID), obtaining comparable scores to state-of-the-art generative methods (FID: 0.0459). We also trained a U-Net for the MRI-based fat quantification task incorporating synthetic datasets. When we used a few real (10 subjects, $~200$ slices) and numerous synthetic samples ($>3000$), fat fraction at specific liver ROIs showed a low bias on data obtained using the same protocol than training data ($0.10\%$ at $\hbox{ROI}_1$, $0.12\%$ at $\hbox{ROI}_2$) and on data acquired with an alternative protocol ($0.14\%$ at $\hbox{ROI}_1$, $0.62\%$ at $\hbox{ROI}_2$). Future work will be to extend PI-LDM to other q-MRI applications.
\end{abstract}


\section{Introduction}
\label{sec:introduction}
\IEEEPARstart{M}{agnetic} resonance imaging (MRI) is a radiation-free imaging modality that allow the acquisition of different contrasts, making it appropriate to visualize various tissue structures. Moreover, in recent years there have been significant advances towards quantitative MRI (q-MRI) applications, which have the potential to facilitate standardized MRI-based measurements, reduce bias, and increase reproducibility \cite{Cheng2012}.

Fat quantification is one of the most validated q-MRI applications \cite{Weingartner2022}. By considering spectral differences between the signals emitted by water and fat protons, it is possible to separate the overall MR signal into water-only and fat-only components. The separated signals can be used to estimate a proton density fat fraction (PDFF) that corresponds to the ratio between fat-only MR signal and the overall MR signal. PDFF maps contain values in the range of 0 to 100\% at each voxel.

PDFF is an extensively validated biomarker for liver malfunctions, such as metabolic associated steatotic liver disease (MASLD), \cite{Eslam2020,Yokoo2018,Idilman2013} and other various pathologies beyond the liver \cite{Idilman2022}. Moreover, PDFF has demonstrated an optimal reproducibility across MR scanners from different manufacturers and varying MR scan protocols and hardware (i.e.: number of coils) \cite{Daude2023,Hernando2017}.

Most water-fat separation techniques require of the magnitude and phase of chemical shift-encoded (CSE)-MR images, which are multi-echo gradient-echo acquisitions. The CSE-MR signal at each echo not only depends on the water and fat signals, but also on variables such as R2* signal decay and magnetic field inhomogeneities ($\phi$), which have non-linear effects on the resulting signal. Therefore, most of state-of-the-art MR water-fat separation methods imply high computational costs to solve a complex-valued non-linear regression from the multi-echo signal samples and may require fine-tunning of parameters \cite{Daude2023}. 

In recent years, several deep learning (DL)-based approaches have been proposed to overcome these drawbacks \cite{Andersson2019,Liu2020,Shih2021,Jafari2021,Meneses2023,Meneses2024,Yang2023}. In particular, some of these DL-based methods have shown promisingly accurate and almost instantaneous PDFF estimations using multi-echo images with fewer echoes than the required by the standard techniques \cite{Meneses2023}, which is usually greater than or equal to six \cite{Yu2007}. 

Nevertheless, a limitation of DL-based water-fat separation methods has been the scarce large databases, due to the specific requirements that the training samples must accomplish. Specifically, DL models usually consider input requirements in terms of the chosen number of echoes, the magnitude-only or complex-valued nature of the input, and specifications of MR scan parameters such as the echo times (TEs) and the main magnetic field strength. As a consequence, it is not straightforward to transfer datasets from one site to another, not only because of data privacy issues, but also because of possible discrepancies on any of the aforementioned variables (i.e.: data format, MR scan parameters, etc.). Moreover, even if large databases are available, it is a significantly time-consuming task to create paired datasets for supervised learning, since reference (and sub-optimal) water-fat separation results must be calculated using some reference state-of-the-art method for each of the CSE-MRI samples.

To alleviate the lack of paired and heterogeneous databases to train DL-based MR water-fat separation methods, we propose a physics-based generative model to synthesize coherent series of CSE-MR images. For this purpose, we developed a physics-informed latent diffusion model (PI-LDM) to synthesize all the variables involved in the signal generation process (i.e., water-only \& fat-only signals, R2* signal decay, field inhomogeneities) to subsequently generate CSE-MR images based on the signal generation model. 

The main contribution of this work is a novel physics-based generative model to simultaneously synthesize CSE-MR images along with their respective parametric maps, that can be easily transferred to be used for augmenting data in other studies. The hypothesis is that PI-LDM synthesized databases will improve, or at least equalize, the performance of DL-based water-fat separation methods to be trained on this data.

\section{Related work}
Previous approaches have explored applications to address the lack of large training databases to train DL methods by generating synthetic data. In this section we will explore some previously proposed data synthesis methods and we will demonstrate how innovative is our approach in the context of q-MRI applications.

\subsection{Physics-based synthetic data generation}
Physics-based data synthesis is an emerging paradigm that can provide huge training data without or with few real data \cite{Yang2023}. Physics-based synthesis is particularly useful in the case of q-MRI, since it is possible to leverage the biophysical models that describe how the MRI signals arise from the underlying tissue parameters to synthetically generate a wide variety of high-quality data \cite{Vasylechko2022}. Then, it is possible to obtain paired synthetic databases since both the MR signals and their corresponding parameters can be generated. A remarkable approach is the work of Vasylechko and coauthors \cite{Vasylechko2024}, who proposed a generative model to synthesize 3D datasets for myelin water fraction mapping, an application in which the data is scarce (i.e., the work of \cite{Vasylechko2024} included MR acquisitions from 6 healthy volunteers) and the reference solutions are significantly unreliable.

\subsection{DL-based generative models}
On the other hand, another relevant approach to address the lack of data are DL-based generative models, which have shown remarkable results in alleviating the lack of medical imaging data \cite{Zhao2023,Khader2023}. In the case of MRI data synthesis, numerous approaches have been proposed to generate multi-contrast MR images from randomly sampled latent spaces or from medical images of different modalities \cite{Yang2023,Dayarathna2023}, and many of them have demonstrated remarkable performance on downstream tasks using DL-based methods trained on synthetic data \cite{Han2020}. For instance, Han and coauthors \cite{Han2020} demonstrated the usefulness of synthetic data for training a classifier to detetc pneumonia from X-ray images.

More recently, Denoising Diffusion Probabilistic Models (DDPM) have shown superior generative performance on computer vision tasks, showing a generative quality comparable to Generative Adversarial Networks (GANs), which used to be the state-of-the-art approach for generative tasks \cite{Han2020}. In contrast to GANs, DDPMs can achieve an improved data distribution coverage and a more stable training process, at a cost of a slower generative process \cite{BondTaylor2022}. DDPMs have been implemented in some MRI-related applications \cite{Dayarathna2023,Khader2023,Kazerouni2023}. A particular application of interest are Latent Diffusion Models (LDM), which are based on solving a reverse diffusion process on the latent space domain created in the middle of an encoder-decoder model \cite{Rombach2021}. LDMs have been successfully applied on medical imaging \cite{MullerFranzes2023,Khader2023} and MRI-based tasks \cite{Pinaya2022,Graham2023}, particularly on brain MRI generation and out-of-distribution detection.

However, none of the previous approaches was specifically designed for q-MRI tasks, where it is necessary to have ground-truth values for task-specific parameters. Moreover, PDFF mapping also requires series of coherent multi-echo images, which would be unfeasible to synthesize using the aforementioned generative models.

\subsection{Combining physics with generative models}
Physics-based generative models leverage the efficiency and realism of AI-based generative models, while enhancing explainability through the incorporation of well-defined principles governing the signal generation process \cite{Yang2023}. Physics-based generative models have shown feasibility in other medical imaging areas such as cryo-electron microscopy generation \cite{Zhong2021}.

In the case of MRI, Jacobs et al. \cite{Jacobs2023} developed a physics-informed deep learning-based method to synthesize multiple brain MRI contrasts from a single ﬁve-minute acquisition. In this work, the proposed GAN-based model maps the MR data to “effective” quantitative parameter maps (PD, T1, and T2) to subsequently use the MR signal model to synthesize four standard contrasts (PD, T1w, T2w, and FLAIR). In this way, the proposed method incorporates, in a data-driven manner via the effective quantitative maps (q*-maps), the imperfections that the signal model does not take into account. The authors' conclusion was that their q*-maps-based approach provides generalizability to unseen contrasts during inference by varying the desired sequence parameters in the signal model. As described in the following sections, our generative model works based on similar principles, since we will design a DL-based model to calculate pseudo-quantitative maps associated to the CSE-MRI signal generation process, to posteriorly reconstruct realistic CSE-MR images using the forward model. 

Another remarkable approach is the InVAErt network proposed by Tong et al. \cite{Tong2024}. In this case, a comprehensive framework for data-driven analysis and synthesis of parametric physical systems was proposed, which uses a deterministic encoder and decoder to represent forward and inverse solution maps, a normalizing flow to capture the probabilistic distribution of a system's outputs, and a variational encoder (VAE) designed to learn a compact latent representation for the lack of bijectivity (i.e., one-to-one function for the complete output domain) between inputs and outputs \cite{Tong2024}. Our generative model will also consider a physics-informed VAE (PI-VAE) but is based on an LDM arhitecture in which a reverse diffusion process is designed to recover PI-VAE's latent space distribution. Moreover, in our case, the PI-VAE is designed to preserve the quality of the reconstructed MR images (i.e., the system's output) instead of the associated parametric maps, which will be determined on an intermediate step by a decoder from the VAE's latent space that can also be retrieved.

\section{Methods}
\subsection{Chemical Shift-Encoded MRI Signal Generation}
Most of the MR water-fat separation techniques assume the following physical model to describe the CSE-MR signal generation process:
\begin{equation}
    \begin{split}
    \mathcal{H}: I(\rho_W,\rho_F,R_2^*,\phi;t) = & e^{-R_2 \cdot t} e^{i2\pi\phi\cdot t} \\
    & \cdot \left[ \rho_W + \rho_F \cdot \sum_{p} \alpha_p e^{i2\pi f_p \cdot t} \right] ,
    \end{split}
    \label{eq:complex-WF}
\end{equation}
where $I$ is the complex-valued CSE-MR signal, $t$ is the echo time, and $f_p$ and $\alpha_p$ are the $P$ peak frequencies of the triglyceride spectrum and their respective relative amplitudes ($\sum\alpha_p=1$). When a multi-peak fat signal model is assumed, $P=6$ peaks are considered with \textit{a priori} known frequencies and amplitudes \cite{Hamilton2011}.

As previously mentioned, the six quantitative parameters that affect the resulting MR signal in eq. \ref{eq:complex-WF} are the complex-valued water and fat proton densities ($\rho_W$, $\rho_F$; two components each), along with the $R_2^*$ signal decay ratio and the off-resonance field ($\phi$). The assumption that each chemical specie has its own magnitude and phase enables the use of complex-valued non-linear least squares methods \cite{Bydder2011}.

However, it is also reasonable to assume that, at an echo time of zero, the initial phase is the same for water and fat, since it is a property of the receiver coil and the used radiofrequency (RF) pulse \cite{Bydder2011}. Considering this assumption, the CSE-MR signal model can be rewritten as:
\begin{equation}
    \begin{split}
    \mathcal{H}: I(\rho_W,\rho_F,R_2^*,\phi,\phi_0;t) = & e^{-R_2 \cdot t} e^{i2\pi\phi\cdot t} e^{i2\pi\phi_0} \\
    &  \cdot \left[ \rho_W + \rho_F \cdot \sum_{p} \alpha_p e^{i2\pi f_p \cdot t} \right] ,
    \end{split}
    \label{eq:mag-WF}
\end{equation}
where $\rho_W\in\mathcal{R}$, $\rho_F\in\mathcal{R}$, and $\phi_0$ is the common initial phase for water and fat. In this case, the number of quantitative parameters is reduced to five.

\section{Physics-Informed Latent Diffusion Model}
Denoising Diffusion Probabilistic Models (DDPMs) are based on gradually denoising a normally distributed variable, which corresponds to learning the reverse process of a ﬁxed Markov Chain of length $T$ \cite{Rombach2021}. The goal is to obtain a denoising auto-encoder, usually a U-Net, capable of removing the noise added at each forward diffusion iteration. In this way, a reverse diffusion process enables to learn the source distribution of the data $p(x)$.

In the case of LDMs, this diffusion process is carried on the latent space domain. Therefore, to implement an LDM, it is necessary to design an autoencoder to obtain a meaningful latent space from CSE-MR images. We propose a Physics-Informed Variational Auto-Encoder (PI-VAE) to reconstruct CSE-MR images whose generalized architecture is comprised by three blocks: an encoder ($\mathcal{E}$), a decoder ($\mathcal{D}$), and a CSE-MR signal model block ($\mathcal{H}$). Thus, a series of multi-echo CSE-MR images acquired at echo times $TE_1,...,TE_N$ can be reconstructed by serially applying all these blocks:
\begin{equation}
    \left[\hat{I}_n\right]_{n=1}^N = \mathcal{H}\left(\mathcal{D}\left(\mathcal{E}\left(\left[I\right]_{n=1}^N\right)\right);TE_1,...TE_N\right)
    \label{eq:PI-VAE}
\end{equation}
The encoder translates the input CSE-MRI data into a latent space $z\in Z$ of reduced dimensions, which is weakly regularized to follow a multi-dimensional Gaussian-like distribution. Then, an LDM with parameters $\Theta$ is trained to approximate the distribution of the latent space domain $Z$, by learning how to iteratively subtract the noise added at each diffusion step $t\in 1,...,T$. New CSE-MRI data $\Tilde{I}_n$ can be generated, at customized echo times, from randomly generated latent spaces $\hat{z}_\Theta$, which are obtained from Gaussian noise $\epsilon$ by solving a $T$-steps reverse diffusion process: 
\begin{equation}
    \left[\Tilde{I}_n\right]_{n=1}^N = \mathcal{H}\left(\mathcal{D}\left(\hat{z}_\Theta(\epsilon;T)\right);TE_1,...,TE_N\right)
    \label{eq:PI-LDM}
\end{equation}
In the following sections we describe our proposed autoencoder and latent diffusion models in more detail.

\subsection{Physics-Informed Variational Autoencoder}
Our proposed Physics-Informed Variational Autoencoder (PI-VAE) is a DL-based parametric model to handle the intractability of the MR water-fat separation task, as well as to learn a low-dimensional embedding of the input CSE-MRI data. Then, the obtained stochastic latent spaces are supposed to encode all the possible q*-maps solutions for each CSE-MRI sample given as input.

\begin{figure*}[h]
\centering
\includegraphics[scale=.44]{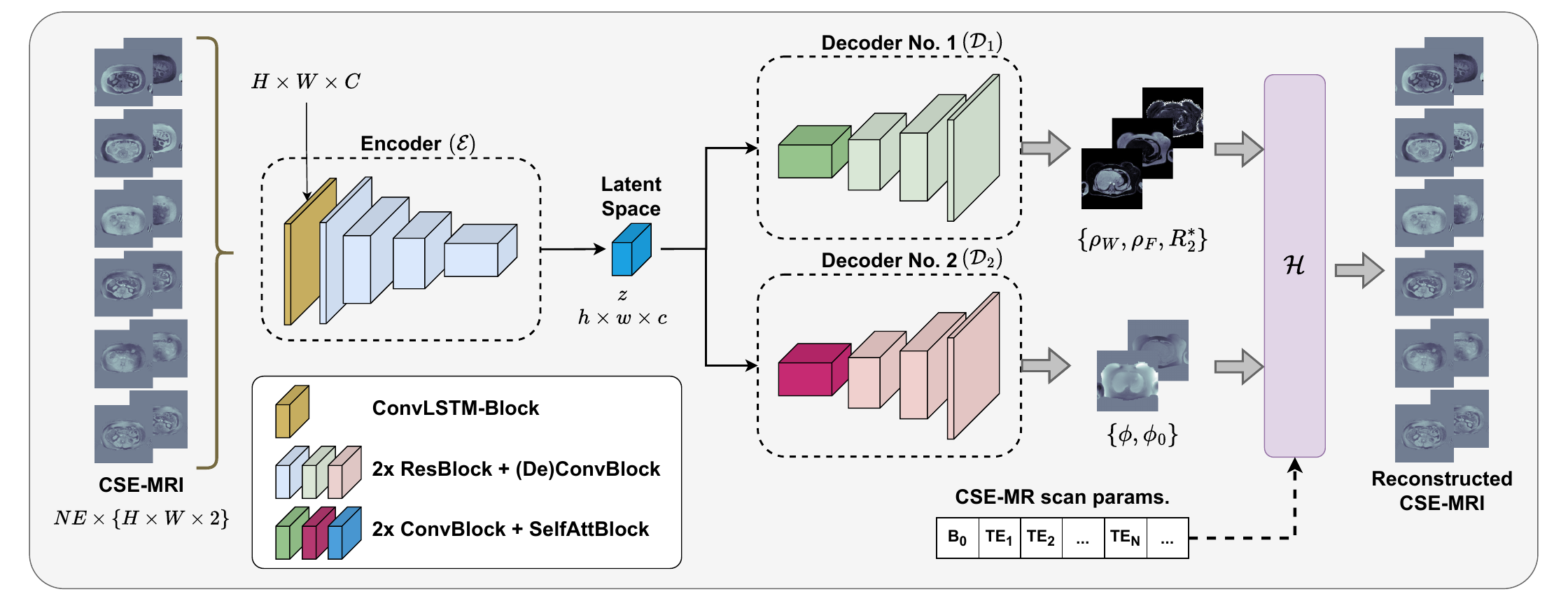}
\caption{Physics-Informed Variational Autoencoder (PI-VAE). The encoder-decoder structure estimates the effective quantitative maps (q*-maps), and an intermediate latent variable ($z$), from CSE-MR images. The CSE-MRI signal model $\mathcal{H}$ is then used to obtain, from the estimated q*-maps, reconstructed CSE-MR images with customized scan parameters (i.e., main magnetic field strength, echo times, etc.).}
\label{fig:VAE}
\end{figure*}

PI-VAE is composed by an attention-based encoder-decoder convolutional neural network (CNN) and a forward model-block (Figure \ref{fig:VAE}).

\subsubsection{The Encoder}
The encoder block ($\mathcal{E}$) is firstly composed of a 2D convolutional long short-term memory (LSTM) layer that processes the multi-echo CSE-MR images into an array of $C=36$ bi-dimensional features. Subsequently, these extracted features are processed by an encoder that outputs a meaningful latent space of reduced dimensionality (based on Figure \ref{fig:VAE}, $h=w=48$, $c=3$). 

\subsubsection{Attention-based Decoder}
The decoder block ($\mathcal{D}$) translates the latent space into five ``effective" quantitative maps (q*-maps), that correspond to each of the parameters involved in the forward model \ref{eq:mag-WF}. The decoder block comprises to separated and independent CNNs: the first one ($\mathcal{D}_1$) calculates the magnitude-based parameters ($\rho_W$, $\rho_F$, $R_2^*$) and the second one ($\mathcal{D}_2$) returns the phase-based q*-maps ($\phi$, $\phi_0$). In the case of $\phi_0$, the output corresponds to the unwrapped phase. Each of the decoders' CNNs has self-attention blocks \cite{Vaswani2017} to process the latent space before upsampling, to improve the non-local estimation of the resulting q*-maps.

\subsubsection{Forward Model Block}
Finally, a forward model operator based on \ref{eq:mag-WF} is used to reconstruct multi-echo CSE-MR images at the same echo-times of the input CSE-MRI samples. \\

Additionally, a discriminative model with a conditional PatchGAN architecture is also considered \cite{Demir2018}. The discriminative model considers two inputs: the odd echoes of the CSE-MR images to be assessed (i.e.: real or reconstructed), and the even echoes of the CSE-MR images given as input to the generator (i.e.: always real during training).

The considered loss function to fit PI-VAE was:
\begin{equation}
    \mathcal{L} = \lambda_R\mathcal{L}_R + \lambda_Q\mathcal{L}_Q + \lambda_D\mathcal{L}_D + \lambda_{KL}\mathcal{L}_{KL}
    \label{eq:loss}
\end{equation}

which has the following components:
\begin{itemize}
    \item Reconstruction loss ($\mathcal{L}_R$): LPIPS perceptual loss \cite{Zhang2018} between the input CSE-MR images ($I_n$) and the VAE reconstruction $\hat{I}_n=\mathcal{H}\left(\mathcal{D}_1\left(\mathcal{E}(I_n)\right);\mathcal{D}_2\left(\mathcal{E}(I_n)\right)\right)$, at each of the $n=1,\ldots,N$ considered echoes.
    \item q*-maps loss ($\mathcal{L}_Q$): Mean absolute error between the q*-maps calculated by the VAE block and the reference q-maps available on the training dataset. To compare against Graph Cuts reference results, $\rho_W$ and $\rho_F$ were considered as complex-valued, each of them with a $\phi_0$ phase.
    \item Adversarial loss ($\mathcal{L}_D$): Conditional discriminator's Wasserstein loss.
    \item KL-regularization ($\mathcal{L}_{KL}$): Kullback-Leibler divergence of the latent space with respect to a standard normal distribution. This loss is designed to give coherence to the resulting latent space.
\end{itemize}
The weights of each loss term were $\lambda_R=1\cdot 10^{-1}$, $\lambda_Q=1\cdot 10^{-2}$, $\lambda_D=1$, and $\lambda_{KL}=1\cdot 5^{-7}$, which were determined by performing numerous experiments and evaluating the reconstruction performance on the validation subset.

Since the encoder's LSTM operator enables input series of variable lengths, we considered a random removal of the last echoes' images during training as a data augmentation step. Therefore, at each training
iteration, CSE-MR images with echoes from three to six were processed along with their respective reference Graph Cuts results, which remained the same despite the augmentation step.

It is important to remark that VAEs can also be designed to be stand-alone generative models by using a strong latent space regularization. This strong regularization enables synthesis from latent spaces that are randomly sampled from a Gaussian distribution. In the case of VAEs used for LDMs, the goal is to partially relax the KL regularization to prioritize the reconstruction quality \cite{Rombach2021}.

\subsection{Latent Diffusion Model}
\begin{figure}[h]
\centering
\includegraphics[scale=.185]{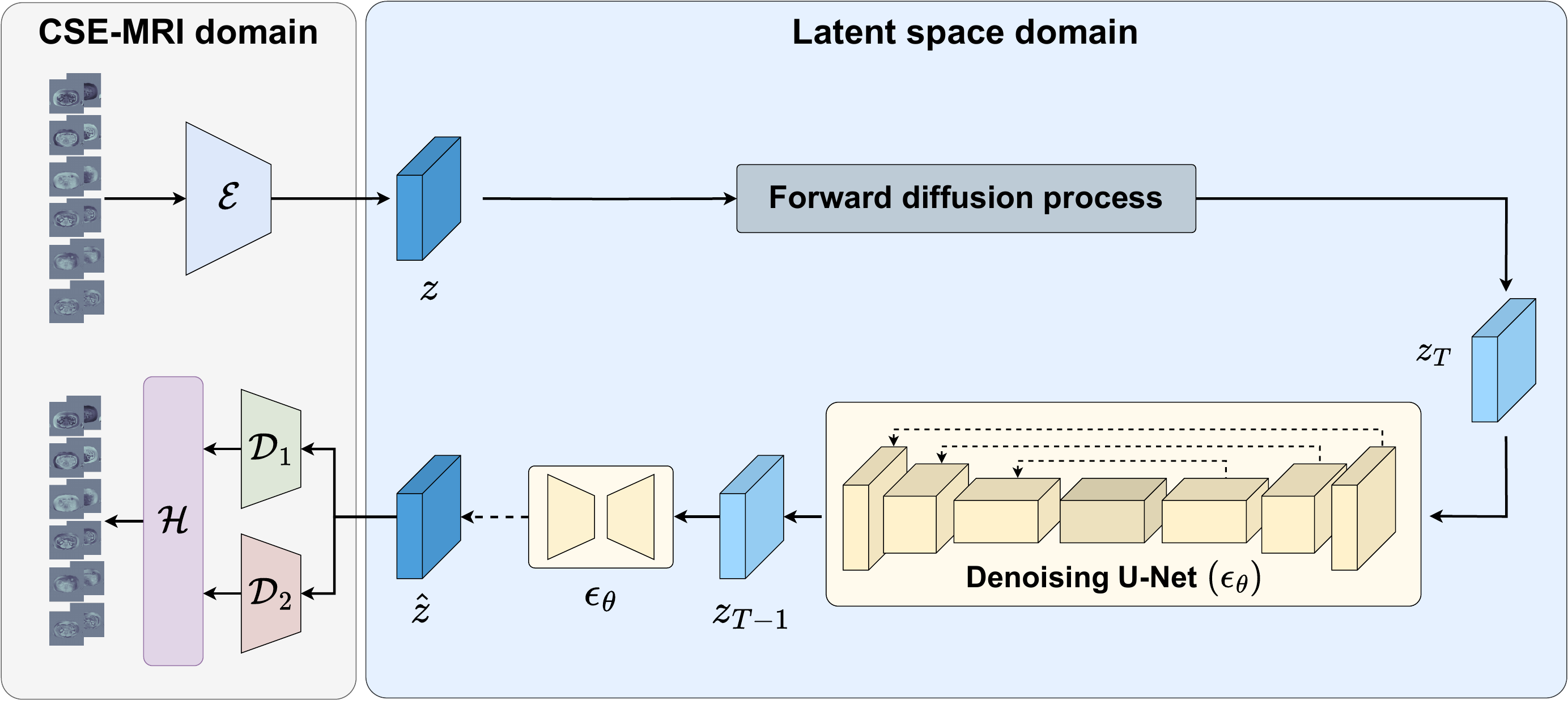}
\caption{Diagram of the proposed Physics-Informed Latent Diffusion Model (PI-LDM).}
\label{fig:LDM}
\end{figure}

Once the previously described PI-VAE was fitted, a denoising U-Net was trained to perform the reverse diffusion process to sample plausible latent spaces that can be successfully decoded by $\mathcal{D}_1$ and $\mathcal{D}_2$ (Figure \ref{fig:LDM}). 

At each iteration during training, the time step $t$ is randomly sampled, in order to cover the complete diffusion process. The noisy samples at each iteration are constructed using the following re-parameterization:
\begin{equation}
    z_t = \sqrt{\bar{\alpha_t}} z_0 + \sqrt{1-\bar{\alpha_t}} \epsilon
\end{equation}
where $\bar{\alpha_t}=\Pi_{i=0}^{t}\alpha_t$, and $\alpha_t = 1-\beta_t$, being $\beta_t$ a linear beta scheduler (or variance schedule) that defines the amount of Gaussian noise added at each step. In this work, the forward diffusion process consisted of $T=500$ time steps. 

Then, the considered loss function was:
\begin{equation}
    L_{DM} = \mathbb{E}_{z,\epsilon\sim\mathcal{N}(0,1),t}\left[ \left|\left| \epsilon-\epsilon_\theta (z_t;t)\right|\right|_2^2 \right]
\end{equation}
where $z=\mathcal{E}(x)$ is the training sample in the latent space domain, $\epsilon$ is the added random Gaussian noise, $z_t$ is a noisy version of the original $z$ obtained after $t$ diffusion steps, and $\epsilon_\theta(z_t;t)$ is the added noise estimated by the denoising U-Net at time-step $t$. Similar to \cite{Rombach2021}, our denoising U-Net considered an auxiliary input that corresponded to the specific diffusion time-step to be solved. Then, considering $z_T=\epsilon$ and iteratively solving the reverse diffusion using:
\begin{equation}
    \hat{z}_{t-1}=\hat{z}_t - \epsilon_\theta(\hat{z}_t,t)~,
    \label{eq:revDiff}
\end{equation}
it is possible to obtain a generated latent space $\hat{z}_\Theta(\epsilon;T)=\hat{z}_0$.

\begin{figure*}[h]
\centering
\includegraphics[scale=.35]{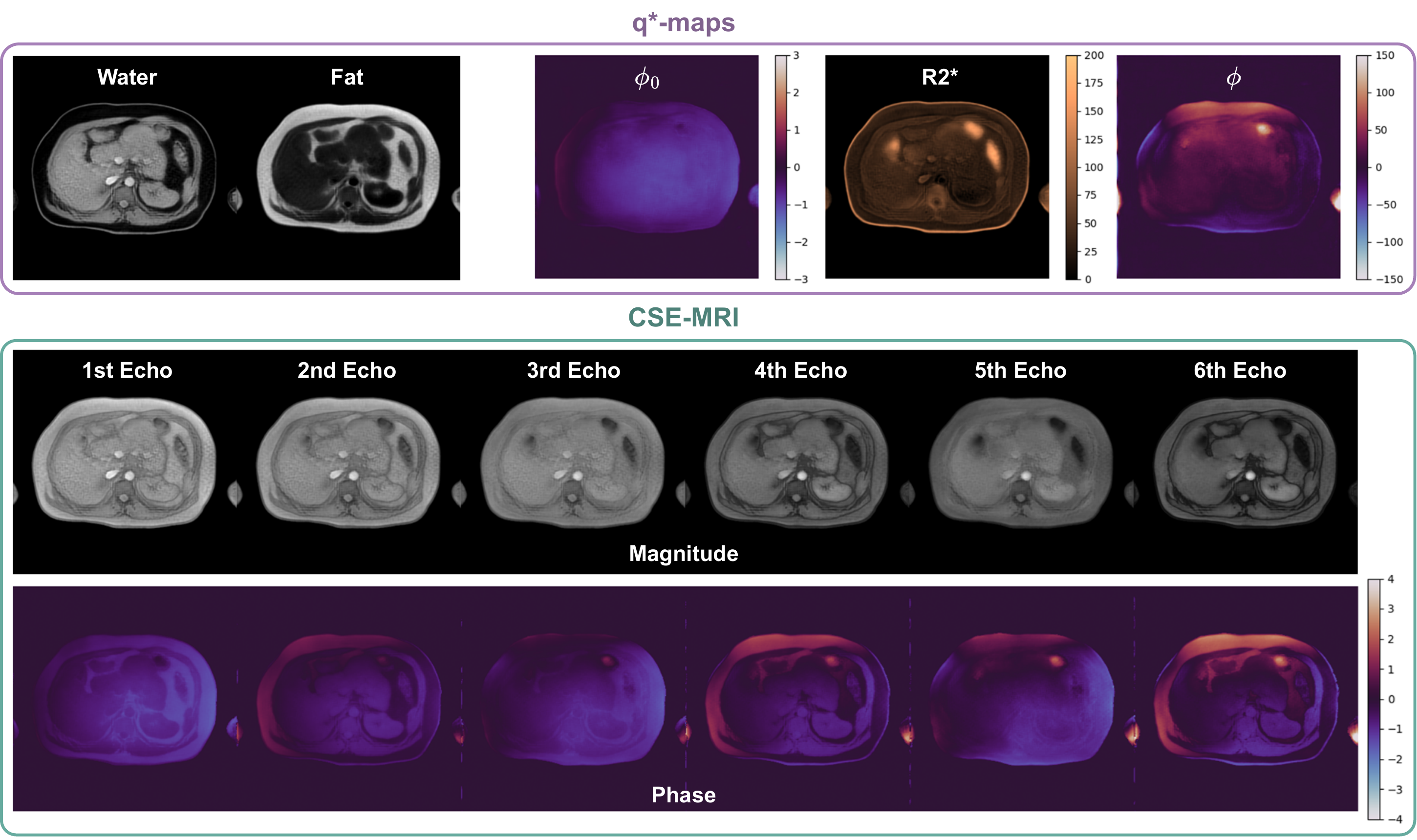}
\caption{PI-LDM synthesized CSE-MR images and their respective q*-maps.}
\label{fig:single}
\end{figure*}

\section{Experiments}
\subsection{Available Data}
Since liver PDFF is a highly validated biomarker, we considered an in-vivo liver CSE-MRI dataset to train and posteriorly validate PI-LDM. This dataset, which has been considered in previous studies \cite{Meneses2023}, included raw-data of multi-slice liver CSE-MR images from 210 subjects (4640 slices), all of them acquired on a single-site (1.5T scanner, Philips, Achieva), and considering a single six-echo protocol with fixed uniformly-spaced TEs. To uniform the size of the MR images, all of them were re-scaled to $384\times384$ by using k-space sub-sampling. Liver CSE-MR images were paired with their respective water/fat separation reference results, including R2* and $\phi$ maps, which were obtained using an iterative Graph Cut-based algorithm \cite{Hernando2010}. Since Graph Cuts solves a complex-valued and non-linear least squares problem, the two proton densities $\rho_W$ and $\rho_F$ are complex-valued.

A fraction of the overall liver MR data was considered to train our generative model (150 subjects, 3330 slices), while a smaller subset was reserved for validation during training (18 subjects, 384 slices). The remaining samples (41 subjects, 926 slices) were separated for the posterior assessment of PI-VAE and other DL-based methods that will be further explained in the following sections.

Additionally, 17 subjects from the testing subset were also scanned using an similar protocol but with different TEs. Specifically this alternative protocol used a TE$_1/\Delta$TE = 1.4/2.2 ms. As explained in the following sections, this data will be useful to assess the performance when variations on the MR scan protocol must be addressed.

\subsection{Generative quality assessment}
To evaluate the quality of PI-LDM's synthesized data, we quantitatively assessed the generated samples' realism. For this purpose, we considered two commonly used metrics to assess the similarity between real-data and synthetic-data distributions \cite{Borji2022}: Fr\'echet Inception Distance (FID) and Maximum Mean Discrepancy (MMD). In the case of FID, the metric relies on a pre-existing classifier (InceptionNet) trained on the ImageNet dataset, and it has demonstrated consistency with human inspection and sensitivity to small changes with respect to the real distribution \cite{Borji2022}. Additionally, similar to the assessment performed by Pinaya et al. \cite{Pinaya2022}, we considered the standard and the multi-scale (MS) Structural Similarity Index Measure (SSIM) between numerous pairs of synthetic samples to assess the diversity of the generated data.

For comparison purposes, we also implemented a stand-alone PI-VAE, which had a stronger KL regularization ($\lambda_{KL}=1\cdot 10^{-6}$) and more down-samplings layers to achieve a smaller latent space (based on Figure \ref{fig:VAE}, $h=w=12$, $c=24$). The liver MRI validation subset was considered as reference; therefore, 384 samples were synthesized by each generative method for this assessment.

\subsection{Evaluation on downstream tasks}
To evaluate the usefulness of our generated data to train DL-based water-fat separation models, we implemented a vanilla U-Net based on previous works \cite{Jafari2021}. This vanilla U-Net considers multi-echo CSE-MR images as input and the output is a 2-channeled array with the water-only and fat-only magnitude images. We performed four experiments that consisted on fitting several U-Net versions using different training data:
\begin{itemize}
    \item \textbf{Real-only dataset}, with the $N_r=3330$ real abdomen slices from the 150 training subjects.
    \item \textbf{Small synthetic-only dataset (S-Synth)} with $N_s=3330$ synthetic abdomen slices (same as real-only dataset).
    \item \textbf{Large synthetic-only dataset (L-Synth)} with $N_s=6660$ synthetic abdomen slices
    \item \textbf{Mixed (real and synthetic) dataset} with real $N_r=200$ real slices from 10 training subjects, and $N_s=3100$ synthetic slices.
\end{itemize}

The resulting U-Nets were tested using the liver MRI testing subset. We measured the mean absolute error (MAE) of the PDFF maps with respect to Graph Cuts estimations. We also estimated the PDFF bias (with respect to Graph Cuts) considering the median of the PDFF at two regions of interest (ROI): the right posterior hepatic lobe (RHL) and the left hepatic lobe (LHL). For each subject, ROIs of $\approx 2~\hbox{cm}^2$ were drawn in the liver parenchyma, at the level of the portal bifurcation, avoiding large vessels and artifacts. Once the ROI locations for each subject were determined, all of them were co-localized on the PDFF maps obtained with each method.

Additionally, we trained two more U-Nets to be used on CSE-MR images obtained with the alternative protocol, with TE$_1/\Delta$TE = 1.4/2.2 ms. The training data for this two versions was:
\begin{itemize}
    \item \textbf{Alternative-TEs synthetic dataset (Alt-Synth)} consisting of $N_s=3330$ synthetic abdomen slices generated considering a TE$_1/\Delta$TE = 1.4/2.2 ms.
    \item \textbf{Alternative-TEs Mixed dataset (Alt-Mixed)} that considered $N_r=206$ real slices from 9 testing subjects scanned using the alternative protocol, and $N_s=3124$ synthetic abdomen slices generated considering a TE$_1/\Delta$TE = 1.4/2.2 ms.
\end{itemize}
These two U-Nets were tested using the CSE-MR images of the remaining 8 testing subjects ($N=226$), considering the same performance metrics of the experiments on the standard protocol (MAE, ROI bias).

\subsection{Implementation details}
All the experiments were conducted on a NVIDIA Quadro RTX 8000 GPU and all DL-based models were implemented using TensorFlow \cite{Abadi2016}. PI-VAE was trained for 200 epochs, using a batch size $BS=2$, learning rate $LR=1\cdot10^{-3}$, and the Adam optimizer. The whole fitting process elapsed 82 hours and 42 minutes. PI-LDM was trained for 300 epochs ($BS=8$, $LR=7\cdot10^{-5}$), elapsing 29 hours. Water-fat separation U-Nets were trained for 100 epochs ($BS=16$, $LR=1\cdot10^{-3}$), with an elapsed time of 10 hours and 40 minutes.

\begin{figure*}[h]
\centering
\includegraphics[scale=.4]{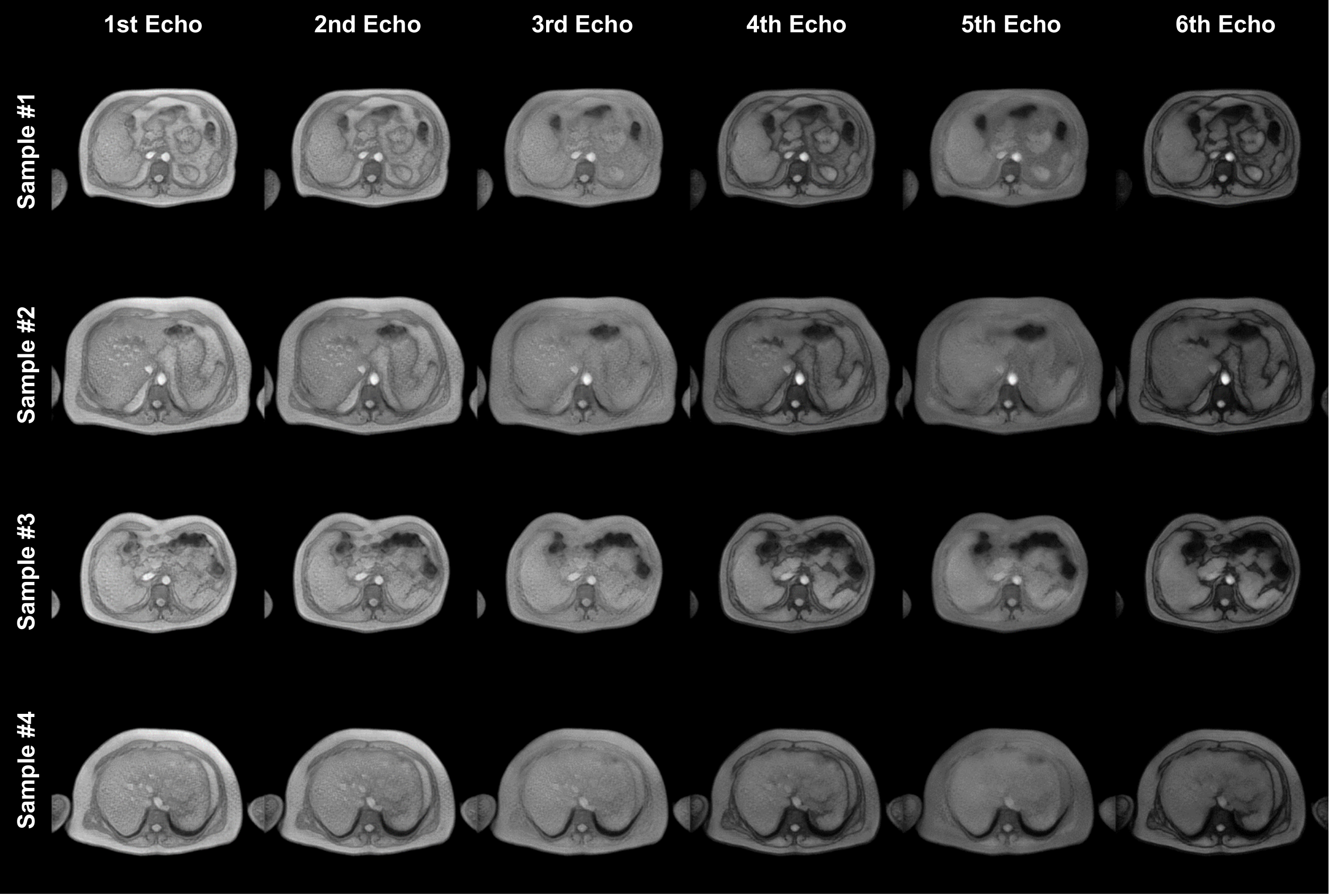}
\caption{Illustration of PI-LDM synthesized CSE-MR images.}
\label{fig:MRgen}
\end{figure*}

\section{Results}
\subsection{Quality of Generated Samples}
Compared to the liver CSE-MRI validation subset, PI-LDM displayed an FID score of 4.81\% and an MMD score of 0.03\%. In terms of data diversity, the observed SSIM and MS-SSIM were 76.15\% and 40.11\%, respectively. On the other hand, stand-alone PI-VAE showed FID and MMD scores of 16.50\% and 0.09\%, respectively. In terms of data heterogeneity, SSIM was 84.14\% and MS-SSIM was 64.29\%. All these metrics are summarized in Table \ref{tab:gen}

\begin{table}[!t]
\caption{Metrics of generative performance, considering the liver CSE-MRI validation subset as reference.}\label{tab:gen}
\centering
\begin{tabular}{|c|c|c|c|c|}
\hline
Model & FID$\downarrow$ & MMD$\downarrow$ & SSIM$\downarrow$ & MS-SSIM$\downarrow$ \\
\hline
PI-VAE & 24.62\% & 0.33\% & 86.56\% & 69.56\% \\
\hline
\textbf{PI-LDM} & \textbf{4.59\%} & \textbf{0.04\%} & \textbf{76.01\%} & \textbf{49.90\%} \\
\hline
\end{tabular}
\end{table}

\begin{figure*}[h]
\centering
\includegraphics[width=\linewidth]{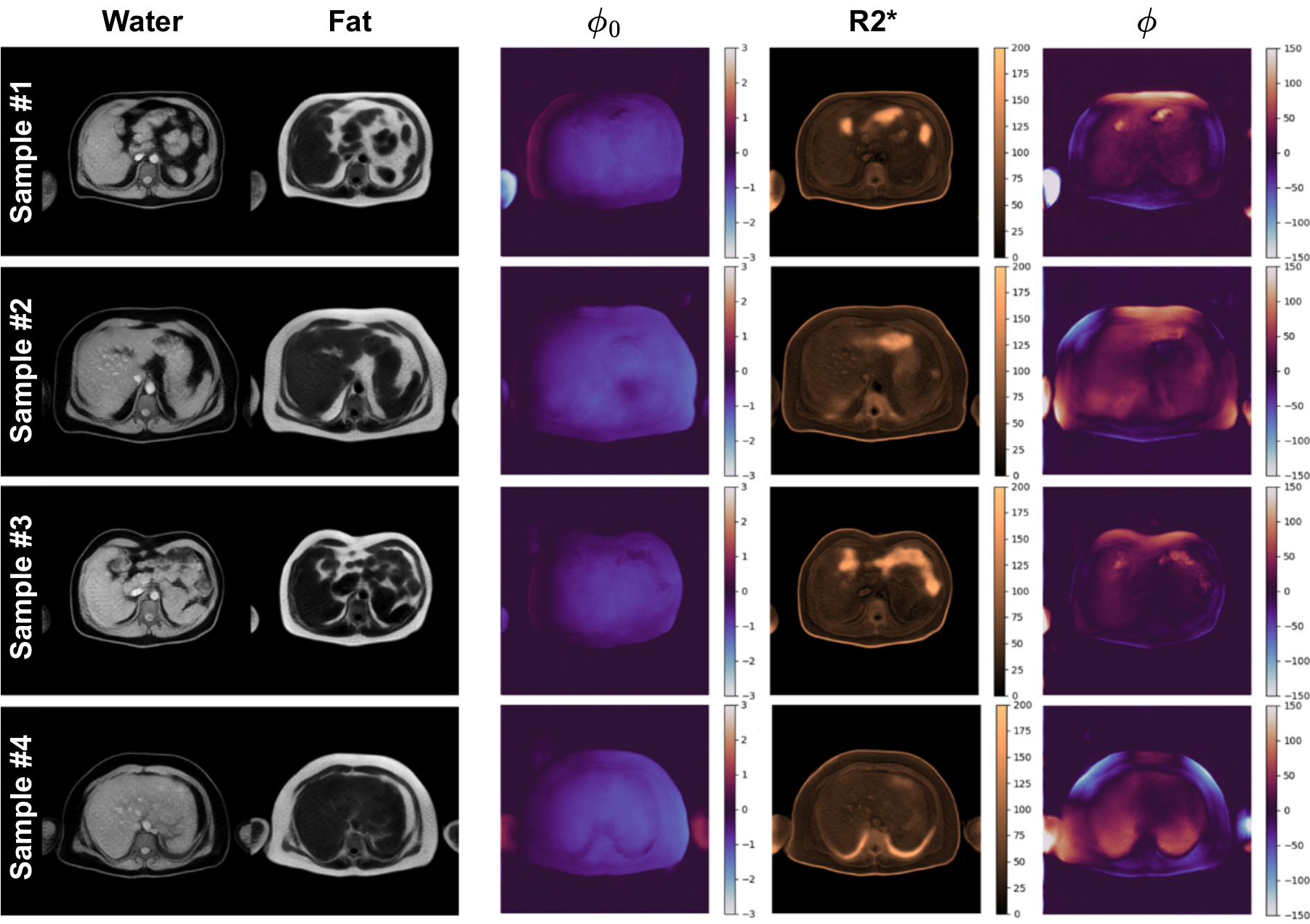}
\caption{Associated q*-maps for each generated CSE-MRI showed in Fig. \ref{fig:MRgen}.}
\label{fig:Qgen}
\end{figure*}

For a visual assessment of PI-LDM generated samples, a synthetically generated CSE-MRI sequence of images, along with their respective q*-maps, is shown in Figure \ref{fig:single}. Additionally, to observe the heterogeneity in terms of geometries and compositions, several generated CSE-MR images are depicted in Figure \ref{fig:MRgen}, and their respective q*-maps are shown in Figure \ref{fig:Qgen}.

\subsection{DL-based Water-Fat Separation}
The first implemented U-Net was trained using the real samples of the liver MRI training subset. In this case, the overall PDFF-MAE ($\pm$ 95\% confidence interval) was $1.33\pm1.08$\%. The observed biases were $-0.50\pm2.74$\% and $-1.39\pm2.70$\% at RHL and LHL, respectively.

In the case of U-Nets trained using synthetic data only, the S-Synth version showed an overall PDFF-MAE of $2.52\pm1.92$\%, and ROI biases of $-1.42\pm2.82$\% and $-1.37\pm2.36$\% at RHL and LHL, respectively. In the case of L-Synth U-Net, the overall PDFF-MAE was $2.37\pm1.84$\%, the ROI bias at RHL was $1.08\pm3.46$\%, and at LHL was $1.86\pm3.23$\%.

Lastly, the U-Net trained with a mixed  database displayed an overall PDFF-MAE of $2.05\pm1.69$\%, and ROI biases of $0.10\pm2.69$\% and $0.12\pm2.39$\% at RHL and LHL, respectively.

All the metrics associated to the PDFF estimation quality of each implemented U-Net are summarized in Table \ref{tab:WFsep}.

\begin{table}[h]
\caption{PDFF metrics of all the implemented U-Net versions. * Last three rows correspond to alternative protocol experiments.}\label{tab:WFsep}
\centering
\begin{tabular}{|c|c|c|c|}
\hline
U-Net Version & MAE (\%) & RHL (\%) & LHL (\%) \\
\hline
Real & $1.33\pm1.08$ & $-0.50\pm2.74$ & $-1.39\pm2.70$ \\
\hline
S-Synth & $2.52\pm1.92$ & $-1.42\pm2.82$ & $-1.37\pm2.36$ \\
\hline
L-Synth & $2.37\pm1.84$ & $1.54\pm3.18$ & $1.37\pm2.70$ \\
\hline
Mixed & $2.05\pm1.69$ & $0.10\pm2.69$ & $0.12\pm2.39$ \\
\hline
\hline
Real* & $1.56\pm1.53$ & $-0.56\pm2.16$ & $-1.62\pm1.75$ \\
\hline
Alt-Synth* & $2.63\pm3.08$ & $2.12\pm3.07$ & $0.75\pm2.87$ \\
\hline 
Alt-Mixed* & $2.33\pm2.18$ & $0.14\pm2.50$ & $0.62\pm2.78$ \\
\hline 
\end{tabular}
\end{table}

\begin{figure}[h]
\centering
\includegraphics[scale=.3]{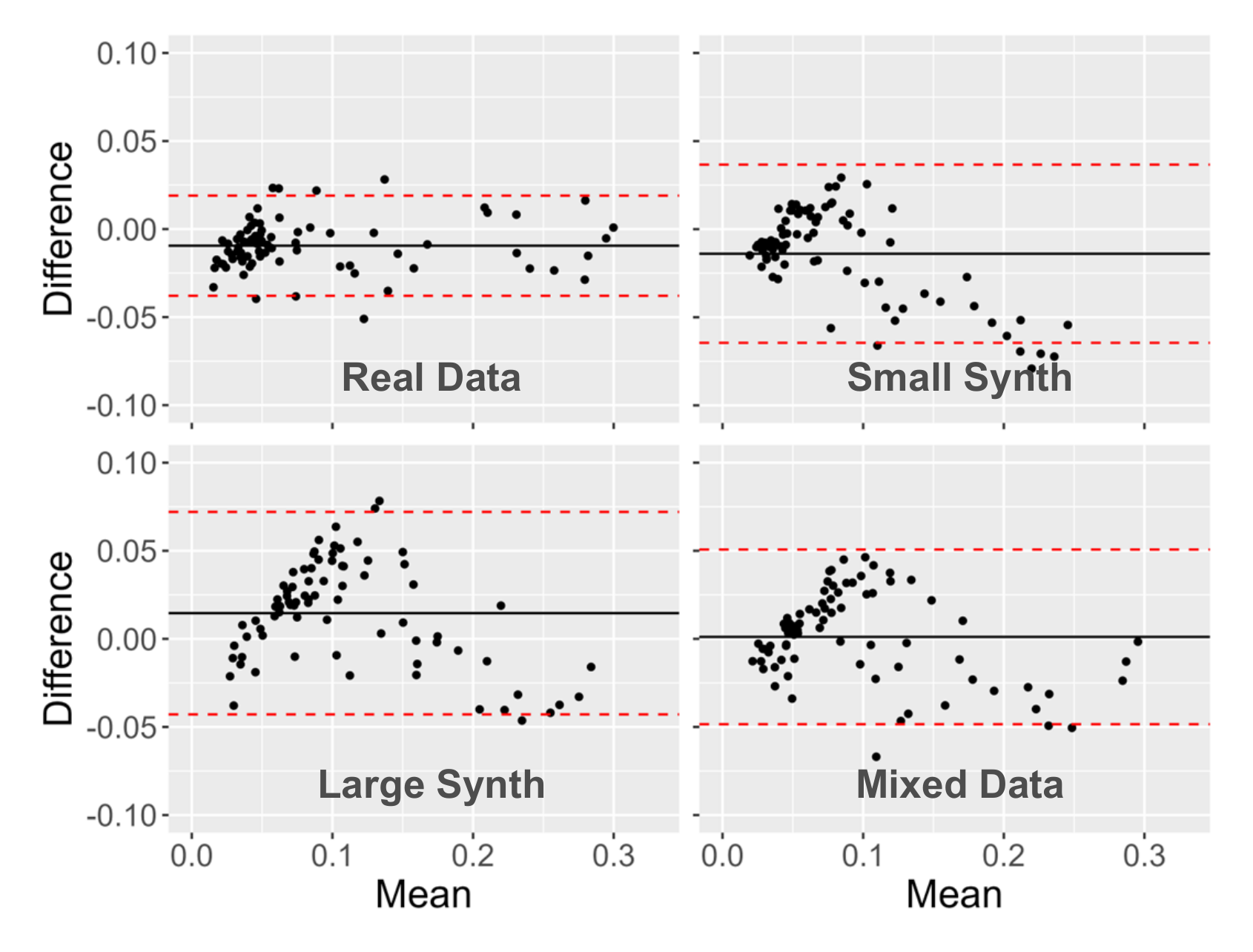}
\caption{Bland-Altman plots considering each U-Net's PDFF estimations at both liver ROIs.}
\label{fig:BA}
\end{figure}

To observe the bias distribution along different fat levels, a Bland-Altman analysis was performed considering the PDFF bias at ROIs. Bland-Altman plots for each implemented U-Net are shown in Figure \ref{fig:BA}.

In the case of the two U-Nets trained to be used on CSE-MR images obtained with the alternative protocol, we observed that the Alt-Synth version showed a MAE of $2.63\pm3.08\%$, and ROI biases of $2.12\pm3.07\%$ at RHL and $0.75\pm2.87\%$ at LHL. In the case of the Alt-Mixed version, the obtained MAE was $2.33\pm2.18\%$, and the ROI biases were $0.14\pm2.50\%$ at RHL and $0.62\pm2.78\%$ at LHL. On the other hand, the real-data U-Net, originally trained with data acquired using the standard protocol, showed a MAE of $1.56\pm1.53\%$, and the observed ROI biases were $-0.56\pm2.16\%$ at RHL and $-1.62\pm1.75\%$ at LHL. All these metrics were also included in Table \ref{tab:WFsep}, although it should be noticed that they were obtained from different datasets.

\begin{figure}[h]
\centering
\includegraphics[scale=.36]{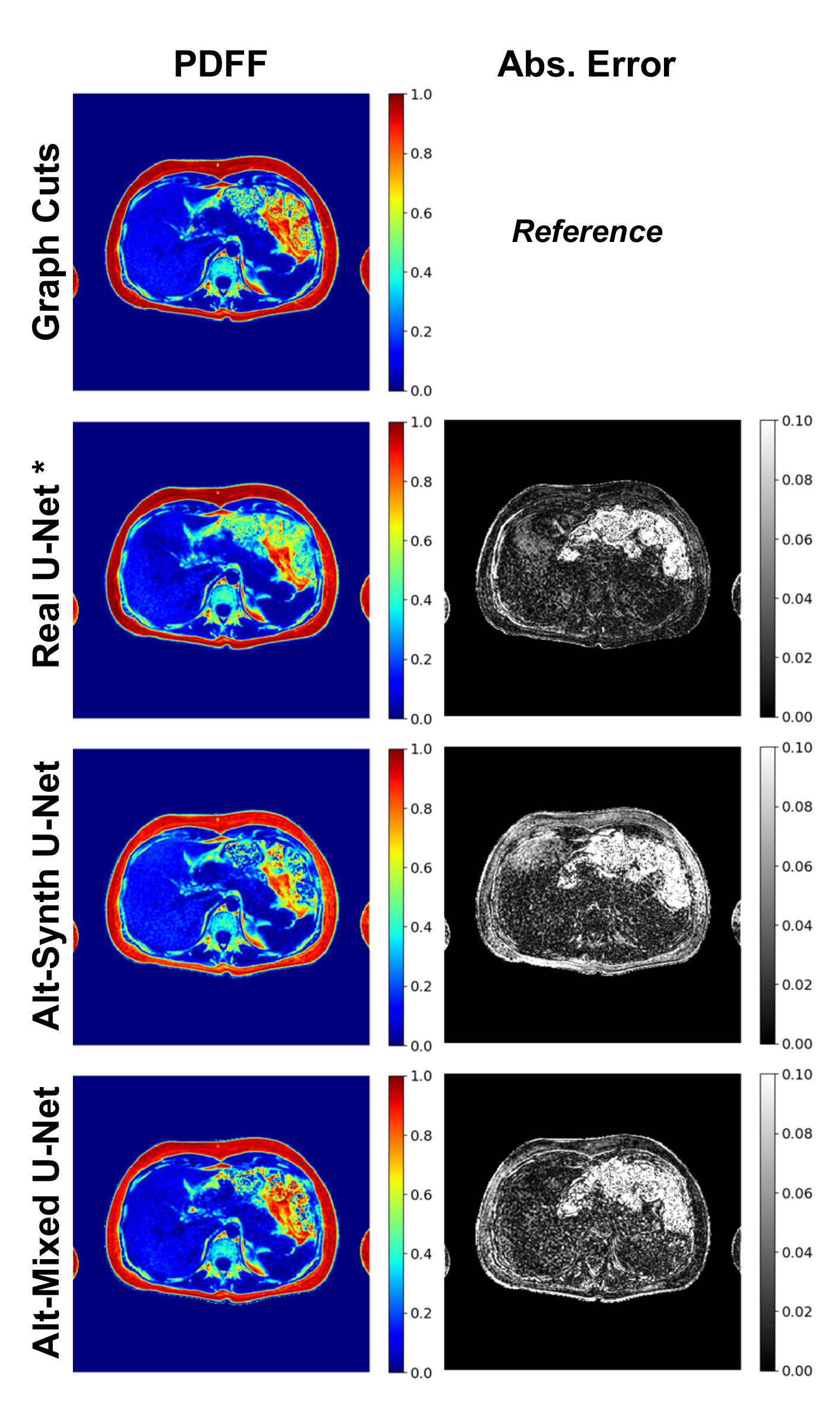}
\caption{PDFF maps obtained for a volunteer scanned using a protocol with TE$_1/\Delta$TE = 1.4/2.2 ms. Each row shows the PDFF and absolute error maps obtained with each implemented U-Net. * The real U-Net was trained with images with different TEs (standard protocol).}
\label{fig:PDFF}
\end{figure}

The resulting PDFF maps for a specific subject of the testing subset that was scanned using the alternative protocol, along with their respective absolute error maps, are shown in Figure \ref{fig:PDFF}.

\section{Discussion}
Synthetic MR image generation have promising potential to be used in various tasks, such as data augmentation, education and training, and discovering new morphologic associations \cite{Khader2023,Kather2022}. Additionally, latent diffusion models have several advantages over diffusion models operating on the image domain, such as a reduction of the computational resources needed during training, the acceleration of the sampling speed, and the operation over a non-pixel-wise and abstract set of features that are extracted in the latent space \cite{Khader2023,Dorjsembe2022}.

In this work, we designed a solution to the lack of large enough and labelled datasets for quantitative MRI applications, particularly addressing the MRI water-fat separation task. Our proposed PI-LDM leverages all the generative power of LDMs, but including a forward model block at the output of the VAE architecture to preserve physics and data consistency, which allows to simultaneously create both CSE-MRI samples and their respective quantitative maps. Additionally, PI-LDM's physics-based synthesis also improves the interpretability and the transparency of the generative process, compared to other DL-based generative approaches, since the compositions of each body region within the image can also be observed.

PI-LDM generated databases demonstrated generative performance metrics comparable to the ones displayed in previously reported works \cite{Pinaya2022}. Moreover, PI-LDM also proved to be useful to train DL-based models for downstream tasks. Specifically, we observed that U-Nets trained for PDFF estimation using PI-LDM's synthesized data showed similar PDFF biases at liver ROIs than similar versions directly trained on real CSE-MRI data. These results suggest that the real training data distribution is being accurately transferred. Moreover, as shown by the Mixed-data U-Net, the data transfer can be significantly enhanced if a few real samples from the original distribution are also considered during training. 


When the alternative protocol with different TEs was considered, we observed that PDFF maps estimated by Real-data U-Net where inaccurate in critical regions of the liver. This result demonstrated that it is not straightforward for a model trained on one protocol to generalize on data obtained using a different protocol. In the case of Alt-Synth U-Net, which was trained purely based on synthetic data with the alternative TEs, PDFF estimation was also sub-optimal. This behavior was expected since there is a distribution shift with respect to CSE-MR images obtained using the original protocol, which were the ones used to train PI-LDM. Contrastingly, Alt-Mixed U-Net's PDFF estimations were significantly de-biased when a few real samples acquired using the alternative protocol were also considered during the U-Net's training. This scenario can be of significant value for many single-center studies that require more training data, as it demonstrates that a relatively small local dataset, in addition to samples from the data distribution learned by PI-LDM, might be sufficient to train accurate enough DL-based models for downstream tasks.

Our proposed PI-LDM has some limitations. Firstly, there was a non-negligible global difference between the PDFF maps estimated with Synth-data models compared to the ones obtained with Real-data U-Net. This difference could be observed in the gap between Synth-data and Real-data U-Nets in terms of the estimated MAEs, although we noticed that this gap was reduced when we considered more synthetic samples. A first possible cause of this behavior is the inaccurate approximation of extreme PDFF values. As shown in Figure \ref{fig:PDFF}, the subcutaneous fat values are underestimated in the PDFF maps obtained from both synthetic data-based U-Nets. Another relevant difference is the amount of noise present on reference Graph Cuts results, which is usually high. It can be observed on PI-LDM generated q*-maps that they tend to be highly de-noised compared to real CSE-MRI and reference q-maps. However, this is not strictly a limitation, since Graph Cuts is a sub-optimal solution for the water-fat separation problem.

Secondly, and related to the aforementioned issues, there are some imperfections regarding both perceptual metrics used in this study (i.e.: LPIPS and FID), since they are based on models pre-trained on natural images. Therefore, although the features extracted by these models are highly correlated to human vision's perceived quality, they might not be sufficiently suitable for this application. At present, quantitative metrics to evaluate generative performance are still a subject of study \cite{Borji2022}.

Thirdly, the forward model that we have considered as the last block of our PI-VAE, is an approximation of the MR signal generation process. Although we have considered the effect of confounding factors such as the R2* signal decay ratio or the magnetic field inhomogeneities, the model dismisses other effects such as the T1 signal decay, or make some assumptions as in the peaks of the fat signal spectrum. PI-LDM's architecture allows modifications on the considered forward model, although the user should notice that the back-propagation during training might be compromised if an excessively complicated MR signal model is considered. 

Future work will be focused in extending our PI-LDM to other q-MRI tasks and in synthesizing three-dimensional data, since we have now ignored potentially useful data dimensions that could improve the performance on generative and downstream tasks \cite{Khader2023}.

\section{Conclusion}
We have proposed a novel physics-based generative model to transfer data distributions for q-MRI tasks, which usually require both MRI data and their corresponding parametric maps. The physics-based nature of our method generates data in a more intuitive and explainable manner than conventional generative methods, with adaptability to scan protocol variations. Since these protocol mismatches are a common issue on multi-center q-MRI studies, we expect that PI-LDM represents a useful tool to transfer data-driven knowledge between centers \cite{Little2023}. We have demonstrated PI-LDM usefulness on the highly-validated MR-based fat quantification task. However, our proposed framework should be extendable to other q-MRI applications in which an analytical model is considered.



\section*{Acknowledgments}
J.P.M. was funded by the National Agency for Research
and Development (ANID)/Scholarship Programme/DOCTORADO BECAS
CHILE/2020—21210665.

\bibliographystyle{IEEEtran}
\bibliography{tmi}

\end{document}